# Surface Compositions Across Pluto and Charon

W.M. Grundy,[1*] R.P. Binzel,[2] B.J. Buratti,[3] J.C. Cook,[4] D.P. Cruikshank,[5] C.M. Dalle Ore,[5,6] A.M. Earle,[2] K. Ennico,[5] C.J.A. Howett,[4] A.W. Lunsford,[7] C.B. Olkin,[4] A.H. Parker,[4] S. Philippe,[8] S. Protopapa,[9] E. Quirico,[8] D.C. Reuter,[7] B. Schmitt,[8] K.N. Singer,[4] A.J. Verbiscer,[10] R.A. Beyer,[5,6] M.W. Buie,[4] A.F. Cheng,[11] D.E. Jennings,[7] I.R. Linscott,[12] J.Wm. Parker,[4] P.M. Schenk,[13] J.R. Spencer,[4] J.A. Stansberry,[14] S.A. Stern,[4] H.B. Throop,[15] C.C.C. Tsang,[4] H.A. Weaver,[11] G.E. Weigle II,[16] L.A. Young,[4] and the New Horizons Science Team.

[1] Lowell Observatory, Flagstaff, AZ 86001, USA.
[2] Massachusetts Institute of Technology, Cambridge, MA 02139, USA.
[3] NASA Jet Propulsion Laboratory, La Cañada Flintridge, CA 91011, USA.
[4] Southwest Research Institute, Boulder, CO 80302, USA.
[5] NASA Ames Research Center, Space Science Division, Moffett Field, CA 94035, USA.
[6] Carl Sagan Center at the SETI Institute, Mountain View, CA 94043, USA.
[7] NASA Goddard Space Flight Center, Greenbelt, MD 20771, USA.
[8] Université Grenoble Alpes, CNRS, IPAG, F-38000 Grenoble, France.
[9] Department of Astronomy, University of Maryland, College Park, MD 20742, USA.
[10] Department of Astronomy, University of Virginia, Charlottesville, VA 22904, USA.
[11] Johns Hopkins University Applied Physics Laboratory, Laurel, MD, 20723, USA.
[12] Stanford University, Stanford CA 94305, USA.
[13] Lunar and Planetary Institute, Houston, TX 77058, USA.
[14] Space Telescope Science Institute, Baltimore, MD 21218 USA.
[15] Planetary Science Institute, Mumbai, India.
[16] Southwest Research Institute, San Antonio, TX 28510, USA.
* Corresponding author. E-mail: w.grundy@lowell.edu

## Abstract

The New Horizons spacecraft mapped colors and infrared spectra across the encounter hemispheres of Pluto and Charon. The volatile ices $CH_4$, CO, and $N_2$, that dominate Pluto's surface, have complicated spatial distributions resulting from sublimation, condensation, and glacial flow acting over seasonal and geological timescales. Pluto's $H_2O$ ice "bedrock" is also mapped, with isolated outcrops occurring in a variety of settings. Pluto's surface exhibits complex regional color diversity associated with its distinct provinces. Charon's color pattern is simpler, dominated by neutral low latitudes and a reddish northern polar region. Charon near infrared spectra reveal highly localized areas with strong $NH_3$ absorption tied to small craters with relatively fresh-appearing impact ejecta.

## Introduction

NASA's New Horizons probe explored the Pluto system in July 2015, returning data from instruments sensitive to electromagnetic radiation from ultraviolet through radio wavelengths, as well as charged particles and dust (1,2). Since the publication of initial results (3), more data have been transmitted to Earth. This paper focuses on spectral and spatial dependence of sunlight reflected from Pluto and Charon in the wavelength range 400 to 2500 nm. These wavelengths are useful for investigating the cryogenic ices prevalent on their surfaces. We restrict our attention to the encounter hemispheres of both bodies since data for the non-encounter hemispheres are as yet incomplete and have lower spatial resolution. Accompanying papers in this issue present results on geology (4), atmospheres (5), the particle environment (6), and small satellites (7).

## Instrument Overview

Data in this paper were chiefly obtained with New Horizons' Ralph instrument (8, Ralph is a name, not an acronym). Ralph consists of a single $f/8.7$ telescope with a 658 mm effective focal length that feeds light to two focal planes: 1) the Multispectral Visible Imaging Camera (MVIC), a visible, near-infrared panchromatic and color imager and 2) the Linear Etalon Imaging Spectral Array (LEISA), a short-wavelength infrared hyperspectral imager. A dichroic beamsplitter transmits IR wavelengths longer than 1.1 µm to LEISA and reflects shorter wavelengths to MVIC.

MVIC is composed of 7 independent CCD arrays on a single substrate. Six large format (5024 × 32 pixel) CCD arrays operate in time delay integration (TDI) mode providing two panchromatic (400 – 975 nm) channels and four color channels: blue ("BLUE", 400 – 550 nm), red ("RED", 540 – 700 nm), near-infrared ("NIR", 780 – 975 nm), and narrow band methane ("CH4", 860 – 910 nm; the filter name is distinguished from the chemical formula by the 4 not being subscript). Using TDI allows very large format images to be obtained as the spacecraft scans the field of view (FOV) rapidly across the scene. A single MVIC pixel is 20 × 20 µrad$^2$ resulting in a 5.7° total FOV in the direction orthogonal to the scan. This width is well matched to the size of Pluto as seen from New Horizons near closest approach. The seventh CCD is a 5024 × 128 element frame transfer panchromatic array operated in staring mode, with a FOV of 5.7° × 0.15°.

LEISA produces spectral maps in the compositionally important 1.25 to 2.5 µm infrared spectral region by imaging a scene through a wedged etalon filter (9) mounted above a 256 × 256 pixel Mercury Cadmium Telluride (HgCdTe) detector array with 62 × 62 µrad$^2$ pixels. LEISA forms a spectral map by scanning the 0.91° × 0.91° FOV across the scene in a push broom fashion. The filter was fabricated such that the wavelength varies along the scan direction. It has two segments: (1) 1.25 to 2.5 µm with an average spectral resolving power of 240 and (2) 2.1 to 2.25 µm with average spectral resolving power of 560.

Supporting observations were obtained with the Long-Range Reconnaissance Imager (LORRI, 10). LORRI's 1024 × 1024 pixel CCD detector has no filter, providing panchromatic response from 350 to 850 nm wavelengths, with a 608 nm pivot wavelength (a measure of effective wavelength independent of the source spectrum, 11). It provides a narrow field of view (0.29°) and high spatial resolution (4.95 µrad pixels). In this paper, the LORRI images are used to provide high resolution geological context imagery and to derive the absolute reflectance of the surface.

# Pluto

New Horizons scanned the LEISA imaging spectrometer across the planet several times on the closest approach date 2015 July 14. We present two LEISA scans, obtained at 9:33 and at 9:48 UTC from ranges of 114,000 km and 102,000 km (12). The resulting spatial scales are 7 and 6 km/pixel, respectively. In combination, the two observations cover the visible disk of Pluto. The data are used to map absorption by various molecules across Pluto's surface, including methane ($CH_4$), nitrogen ($N_2$), carbon monoxide (CO), and water ($H_2O$) ices, revealing that these ices have complex and distinct spatial distributions as described in the next two subsections.

## Pluto's volatile ices

$N_2$, CO, and $CH_4$ ices are all volatile at Pluto's surface temperatures of 35-50 K (13,14). They support Pluto's atmosphere via vapor pressure equilibrium and participate in Pluto's seasonal cycles (15). Of the three, $N_2$ has the highest vapor pressure and thus dominates the lower atmosphere, while $CH_4$ is the least volatile, with a vapor pressure a thousandth that of $N_2$ (13). $N_2$, CO, and $CH_4$ ices are all soluble in one another to varying degrees, so on Pluto's surface, the three ices are likely mixed to some extent at the molecular level (16,17,18). The volatility contrasts and complex thermodynamic behaviors of ice mixtures are expected to produce distinct spatial distributions of these ices across Pluto's surface as functions of season, heliocentric distance, latitude, altitude, local slope, substrate albedo, and thermal properties. LEISA data reveal complex distributions of the volatile ices as shown in Fig. 1. Brighter colors correspond to greater absorption by each ice, but the scale is arbitrary, so only relative variations are meaningful in this context. The geological context is shown by overlaying each colored absorption map on the higher-resolution LORRI base map (4) in the bottom part of each panel.

Methane ice's numerous absorption bands dominate Pluto's near-infrared spectrum. Fig. 1A shows absorption by $CH_4$ ice at 1.3-1.4 µm to be widely distributed across the planet's surface. The $CH_4$ absorption is especially strong in the bright, heart-shaped region informally known (19) as Tombaugh Regio (TR), in Tartarus Dorsa to the east, in the high northern latitudes of Lowell Regio, and in the sliver of the southern winter hemisphere visible south of Cthulhu Regio. $CH_4$ absorption appears relatively uniform across the thousand km wide icy plain of Sputnik Planum (SP), the western half of TR. At northern mid-latitudes, the $CH_4$ distribution is much more patchy, evidently influenced by topographic features (see Fig. S4). Many craters show strong $CH_4$ absorption on their rims but not on their floors, although there is some variability to this pattern. Figs. 1A shows the floors of Burney and Kowal craters having some $CH_4$ absorption whereas those of Giclas and Drake craters look more depleted (see also Fig. S4).

In eastern TR, the region around Pulfrich crater has conspicuously little $CH_4$ absorption. Other areas with weak methane absorption include parts of al-Idrisi and Baré Montes west of SP and the low-albedo equatorial regions Cthulhu Regio and Krun Macula, although a few crater rims and the peaks of a mountainous ridge within Cthulhu Regio do show strong $CH_4$ absorption.

$N_2$ ice was first identified on Pluto from its weak absorption band at 2.15 µm (20). This band has an absorption coefficient ~$10^5$ times less than that of $CH_4$ at similar wavelengths, so the fact that it could be detected at all suggested $N_2$ could be the dominant ice on the surface of the planet. Fig. 1B shows a map of $N_2$ ice absorption from LEISA data. Relatively little absorption is seen at low latitudes, except for SP, where $N_2$ absorption is strong. As with the $CH_4$ absorption, $N_2$ absorption is patchy in northern mid-latitudes, but the spatial distribution is quite distinct from that of $CH_4$. $N_2$ absorption appears strongest on many crater floors, notably those of Burney, Safronov, Kowal, and Drake craters, consistent with topographic control (Figs. 1B and S4). Little $N_2$ absorption is seen in Lowell Regio, possibly related to seasonal sublimation, since high northern latitudes have been exposed to continuous sunlight since the late 1980s (21), but substantial path-lengths are required to produce observable $N_2$ absorption (e.g., 20,22), so lack of absorption does not necessarily exclude its presence. A texture that produces short optical path lengths through the $N_2$ ice could also make it undetectable.

CO ice has absorption bands at 1.58 and 2.35 µm (20,22). Since the 2.35 µm CO band is entangled with adjacent strong $CH_4$ bands, we constructed a CO map using the more isolated 1.58 µm band. This band is very narrow and shallow, producing a noisy map. To help overcome the noise it was spatially binned to 24 × 24 km pixels. The most salient feature in the CO map (Fig. 1C) is greater absorption in Sputnik Planum, most prominently to the south of ~40° N latitude.

SP stands out as the one region of Pluto's encounter hemisphere where all three volatile ices coexist. This region has been interpreted as a cold trap where volatile ices have accumulated in a topographic low, possibly originating as an impact basin (4). The uncratered and thus young surface of SP is apparently refreshed by glacial flow of volatile ices, possibly driven by convective overturning (4). The absorptions of Pluto's two most volatile ices $N_2$ and CO are especially prominent south and east of a line running roughly from Zheng-He Montes to the southern part of Cousteau Rupes. The greater absorption by $N_2$ and CO ices in the core of SP coincides with higher albedos and possibly elevations, perhaps indicating the area of most active or recent convective recycling.

## Pluto's less volatile surface materials

Water ice, heavier hydrocarbons, and other materials had long been sought on Pluto. Absorptions of $H_2O$ and $CO_2$ ices are readily apparent in spectra of Neptune's largest moon Triton (23,24,25), considered an analog for Pluto. Pluto's stronger $CH_4$ absorptions frustrated the unambiguous detection of $H_2O$ from Earth-based observations (e.g., 26). $CO_2$ ice's narrow absorptions have never been reported in remote observations of Pluto, and New Horizons LEISA observations have produced no unambiguous detection of exposed $CO_2$ ice either.

LORRI images of Pluto show mountain ranges bordering SP (3,4).  These mountains, some as high as several km, could not be constructed of the volatile ices $N_2$, $CH_4$, and CO and still endure for geological timescales (27,28).  $H_2O$ ice is the most cosmochemically abundant durable material consistent with Pluto's origins and likely internal structure (e.g., 29).

The broad nature of $H_2O$ ice absorption bands and the plethora of strong $CH_4$ bands make mapping Pluto's $H_2O$ difficult with simple ratios or equivalent widths.  Instead we computed the linear correlation coefficient with an $H_2O$ ice template spectrum (Fig. 2).  The highest correlations are in the vicinity of Pulfrich crater in east TR, and also along Virgil Fossa.  In MVIC enhanced color images, the water-rich region in Virgil Fossa appears distinctly reddish/orange in color (see next section).  High $H_2O$ spectral correlations are seen in several regions in Viking Terra and Baré Montes with similarly reddish/orange coloration in the enhanced MVIC color images.  In contrast, the $H_2O$-rich region around Pulfrich crater looks more neutral in the color images.  Other montes including al-Idrisi, Hillary, and Zheng-He have lower correlation values, but when their spectra are compared with more $CH_4$-dominated spectra like 'a' and 'e', they show clear evidence for water ice via enhanced absorption at 1.5 and 2.0 µm (see Figs. 3, S5).  Localized $H_2O$-rich regions in these areas tend to correspond to valleys between individual mountain peaks or topographic lows as in the core of al-Idrisi Montes, rather than the summits of the mountains.

Cthulhu Regio shows some correlation with the water ice template spectrum, especially toward the west and along the northern and southern flanks of the regio, but Cthulhu's $H_2O$ absorptions at 1.5 and 2.0 µm are relatively shallow.  An absorption around 2.3 µm is probably indicative of hydrocarbons heavier than $CH_4$.  The occurrence of heavier hydrocarbons in Cthulhu Regio is consistent with ground-based observations suggesting that ethane ice absorptions are most prominent at those longitudes (30), though additional hydrocarbons are also likely to be contributing to the absorption in that wavelength region.   A region toward the east of Cthulhu, near the equator, shows little evidence of $H_2O$ absorption and could represent the spot richest in Pluto tholins of all the encounter hemisphere.

## Pluto colors

New Horizons MVIC obtained color images of Pluto on multiple epochs.  We present an observation obtained on 2015 July 14 11:11 UTC, about 40 minutes prior to closest approach, from a range of 35,000 km.  The resulting spatial scale was 700 m/pixel, the best color spatial resolution so far returned to Earth.  Fig. 4A shows an "extended" color view of this data set in which MVIC's BLUE, RED, and NIR filter images are displayed in the blue, green, and red channels, respectively.

Color ratios remove illumination effects and highlight color variability as shown in Fig. 4B.  RED/BLUE and NIR/RED ratios both vary by over a factor of two across Pluto's encounter hemisphere.  Most of this variation is distributed along an axis from blue/neutral colors in the lower left to much redder colors in the upper right, but various clumps and deviations from this axis are indicative of additional subtleties.

Pluto's color diversity is further explored via principal component analysis (PCA), projecting brightnesses in the four MVIC filters into an orthogonal basis set where each dimension successively accounts for the maximum amount of remaining variance. The first principal component (PC1) corresponds to overall brightness across the scene. PC1 accounts for 98.8% of the variance of the MVIC color data, mostly due to illumination geometry and to Pluto's extreme albedo variations (next section). Principal components 2, 3, and 4 account for 1%, 0.12%, and 0.05% of the total variance in the full MVIC data set, respectively. The coherent spatial patterns seen in all three are indicative of distinctly colored provinces across Pluto's surface. Fig. 4, Panels C, D, E, and F show the four principal component images along with the eigenvectors. Panel G combines the principal component images, showing many distinct color units.

We used the narrow MVIC CH4 filter in conjunction with RED and NIR filters to compute a $CH_4$ equivalent width map (Fig. 5) and a color slope map (Fig. S3, details in supplementary text). A key distinction between this and the LEISA $CH_4$ map in Fig. 1 is that they probe two different $CH_4$ ice bands. The 0.89 µm band targeted by the MVIC CH4 filter has a peak absorption coefficient roughly an order of magnitude below that of the 1.3-1.4 µm band complex being mapped in Fig. 1 (31). Consequently, Fig. 5 is mapping greater path lengths in $CH_4$ ice, and thus areas that are especially rich in $CH_4$ ice and/or have especially large particle sizes. The distribution is broadly similar to the LEISA $CH_4$ map, but there are differences. Regions standing out for their strong 0.89 µm band absorption include the bladed terrain of Tartarus Dorsa and low latitude bands flanking Cthulhu Regio. The much higher spatial resolution of the MVIC observation makes it possible to see subtle variations in $CH_4$ absorption within SP. The north, west, and southwest margins of SP show stronger $CH_4$ absorption. In the core of SP, where absorptions of the more volatile ices $N_2$ and CO are more prominent, the boundaries between the polygonal convection cells (described in 1,4) show less $CH_4$ absorption.

When Pluto's known atmospheric gases ($N_2$, $CH_4$, are CO), are exposed to energetic photons or charged particles, chemical reactions produce more complex radicals and molecules that are generally non-volatile at Pluto's surface temperatures (32,33,34). Similar photolytic and radiolytic processing occurs in these same molecules condensed as ices (35,36). Pluto's present-day atmosphere is opaque to Ly-alpha solar ultraviolet (21), so photochemical products are mostly produced in the atmosphere, condense as haze particles, and eventually settle to the surface. Since in the present epoch Ly-alpha does not reach the surface, ices on the surface are currently primarily affected by interstellar pickup ions, galactic rays, and their spallation products from their interactions with the atmosphere. Cosmic rays can induce chemical changes at depths exceeding one meter into the surface (37,38).

Laboratory simulations of radiolysis of a Pluto ice mixture at T~15K (39) forms refractory residues with colors resembling some of the colors on Pluto. Chemical analysis of this material shows atomic ratios N/C~0.9 and O/C~0.2, indicating that the 1.2 keV electrons used in the experiments dissociate the $N_2$ molecule, allowing the N atoms to react with other atoms and molecular fragments. The residue contains urea, alcohols, carboxylic acids, ketones, aldehydes, amines, and nitriles. A substantial aromatic component is found in two-step laser desorption mass spectroscopy, with mass peaks throughout the range ~50-250 dalton.

During any putative epoch when Pluto's atmosphere collapses, it would not shield the surface from ultraviolet photons and solar wind particles as it does now. These would then reach the surface and directly contribute to its chemical evolution. The production of the colored ice residue in the laboratory with low energy electrons occurs in a matter of hours with an electron fluence of ~$10^{17}$/mm$^2$. Charged particles and scattered Ly-alpha can arrive from all directions, so in the absence of an atmosphere, coloration could arise as quickly as a few years, even on unilluminated surface regions, much faster than the ~40 Kyr timescale for tholin haze deposition from Pluto's atmosphere (5).

## Pluto albedos

Four major global albedo units are evident in New Horizons Pluto images: low-albedo equatorial regions exemplified by Cthulhu Regio and Krun Macula, the northern summer polar region Lowell Regio, a sliver of southern winter hemisphere, and the high-albedo TR. TR's albedo is similar to that of Triton (40). Pluto's dark equatorial regions have albedos similar to some outer Solar System moons that are rich in carbonaceous or organic material, such as the saturnian moons Hyperion and Phoebe, and the uranian moon Umbriel, although it is not as dark as the low-albedo hemisphere of Iapetus (41).

In planetary surface images, intensity differences are mostly due to illumination and observing geometry. A photometric function is needed to obtain quantitative measurements of normal albedo (brightness for incident, emission, and solar phase angles all equal to 0°). Fig. 6A is a global map of normal albedo from LORRI images, using a photometric function in which 30% of the reflected photons obey Lambertian scattering, while the rest follow a single-scattering lunar function (see supplementary text). This function is similar to those found for the icy moons of Saturn (41). We also accounted for the 0.04 magnitude opposition surge for the R-filter below 0.10° (42,43).

This albedo map illustrates the quantitative differences in albedo for regions characterized by the distinct combinations of volatile ices and colors seen in preceding figures. Albedo and composition can interact in complex ways: high albedo regions that absorb less sunlight tend to become sites of volatile ice deposition, while low albedo regions can absorb much more sunlight, driving sublimation of volatiles and reaching higher temperatures. Deposition of volatile ices can raise the albedos of regions if they are configured into textures that scatter light, and the texture of mixed volatile ices can change as a result of annealing or sintering, or temperature changes leading to phase transitions or fracturing.

# Charon

## Charon colors

Fig. 7A shows the highest spatial resolution MVIC color observation of Charon from New

Horizons, with a spatial scale of 1.5 km/pixel.  The spacecraft recorded this scan on 2015 July 14 10:42 UTC, about 70 minutes before closest approach, from a range of 74,000 km.  As previously known from Earth-based observations, Charon's surface color is generally neutral (44,45).  New Horizons data reveal a large scale exception with Mordor Macula, the northern polar region, being distinctly red.  The red coloration begins to appear northward of about 45°, as measured by NIR/BLUE and NIR/RED color ratios (Fig. 7B).  In addition to this large scale feature, there are a variety of local color variations.  Craters and other features complicate or interrupt the trend toward redder coloration at high latitudes, such as Dorothy Gale crater, which is less red than the local latitude trend, and Vader crater, which is more red.  Lower latitude color variations include the ejecta of Nasreddin crater being bluer than surrounding terrain, and Gallifrey Macula redder.  North and south of the tectonic belt extending across Charon's encounter hemisphere, colors are similar, but the smoother plains of Vulcan Planum show less color diversity.

Charon MVIC color ratios (Fig. 7C) show a simpler distribution of colors than seen on Pluto.  The bulk of the surface is spectrally neutral, with a mixing trend toward the redder colors at high latitudes.  Principal component analysis of the four colors corroborates this simple color distribution.  As before, PC1 (Fig. 7D) maps brightness across the scene, controlled by albedo and illumination geometry, accounting for 97.3% of the observed variance.  PC2 corresponds to the reddish polar coloration (Fig. 7E), albeit inverted so the pole looks dark.  It accounts for 2.7% of the observed variance, greater than for Pluto's PC2.  Charon's PC3 and PC4 show little coherent structure (Panels F and G), apparently responding primarily to noise.  They account for only 0.03% and 0.02%, respectively, of the variance, much less than their counterparts on Pluto.

## Charon spectral characteristics

New Horizons observed Charon with LEISA from a range of 82,000 km on 2015 July 14 10:30 UTC, at a spatial scale of 5 km/pixel.  The data confirm that Charon's encounter hemisphere is composed predominantly of water ice, as first identified in the mid-1980s (46).  Earth-based observations had also showed that Charon's water ice was at least partially in the crystalline phase, as indicated by the 1.65 μm band, and that the water absorption was seen at all longitudes as Charon rotated (47,48,49).  LEISA observations confirm that water ice is everywhere on Charon's encounter hemisphere, with the 1.5, 1.65, and 2 μm bands being evident in all of the example spectra in Fig. 8.

Spectral observations also revealed an absorption band around 2.22 μm, attributed to ammonia hydrates (47,48,49).  Subsequent studies (50,51) showed that the band varies with sub-observer longitude as Charon rotates.  LEISA observations now show that the ammonia absorption is distributed across Charon's encounter hemisphere at a low level, with local concentrations associated with a few of Charon's bright rayed craters.  Organa crater in the northern hemisphere is the best example.  The crater is about 5 km across, and is thus not resolved by the LEISA pixels.  The $NH_3$ signature appears to be associated with the crater plus some, but not all, of the ejecta blanket (see Figs. 8C, S6).  According to laboratory studies, ammonia ice is destroyed by UV photons and cosmic rays (52,53). From fluxes in Charon's environment (38) the timescale for radiolytic destruction of Charon's $NH_3$ was estimated to be of order $10^7$ years (50), implying that these deposits are relatively recent.

LEISA spectra of Mordor Macula do not reveal distinguishing spectral features coinciding with the red coloration, apart from subtle differences in continuum slope toward the shorter wavelength end of LEISA's spectral range. The reddish colorant may be too thin to produce stronger features at near-infrared wavelengths, or may simply lack distinct absorption bands at LEISA wavelengths.

## Discussion

Various patterns emerge from the observations. Latitude-dependent distributions of materials were expected from seasonal volatile transport processes (54), and indeed, the LEISA and MVIC data confirm a number of distinct latitude zones, especially in the western half of the encounter hemisphere. Pluto's equatorial latitudes feature regions that are strikingly dark and red at visible wavelengths, typified on the encounter hemisphere by Cthulhu Regio and Krun Macula. These provinces are much less dark at infrared wavelengths, and in many areas they show weak 1.5 and 2 µm features of $H_2O$ ice, along with absorptions by hydrocarbons around 2.3 µm (Fig. 3). A possible scenario is that these regions are ancient, heavily cratered landscapes where tholins and other inert materials have accumulated over geological timescales. Flanking the dark equatorial belt to both north and south are higher albedo regions rich in $CH_4$ ice (Figs. 1A and 5). As the least volatile of Pluto's volatile ices, it should be the first to condense and last to sublimate away, consistent with its proximity to the volatile-depleted maculae. The $CH_4$ is most prominent in topographically high regions such as ridges and crater rims, and $CH_4$ can even be found in a few isolated high-altitude regions within the maculae. The bladed terrain of Tartarus Dorsa is especially $CH_4$-rich (Figs. 1A and 5), and could result from many seasonal cycles of $CH_4$ accumulation on elevated low-latitude regions. At northern latitudes above about 35° N, more volatile $N_2$ ice begins to appear, favoring topographic lows where the surface pressure is higher (Fig. 1B). Still further north, $N_2$ and CO absorptions are weak in Lowell Regio (Figs. 1B and 1C), while $CH_4$ absorption continues right up to the pole (Figs. 1A and 5). This high albedo region has been described as a "polar cap," although the lack of prominent $N_2$ and CO ice absorptions makes that term seem poorly suited to describe a summer pole comparatively depleted in Pluto's more volatile ices.

This latitude-dependent distribution of Pluto's surface materials is dramatically interrupted in Tombaugh Regio. The western half of TR is Sputnik Planum, a deep basin hosting a unique, youthful surface morphology described in detail in (4). The spectral signatures of $N_2$, $CH_4$, and CO ices are all present in this region, with the absorptions of the more volatile $N_2$ and CO ices being especially prominent in the southeastern part of SP, below the southern limit for latitudes experiencing the "midnight sun" during the current epoch (21). But this pattern does not seem to be purely governed by climatic factors, since the boundary curves from around Zheng-He Montes in the southwest to Columbia Colles in the northeast. Another potential explanation could involve bulk glacial flow of ices to the northwest, with ablation of the more volatile $N_2$ and CO from the surface of the flow. Alternatively, the locus of most active convection could migrate around within SP, with less active regions showing less absorption by the volatile ices. It is also possible to interpret the reduced volatile ice absorption toward SP's northwest flank as being due to evolution of the surface texture alone, with no change in bulk composition. A reduction in particle size or an increase in scattering could account for the reduced absorption toward that flank. Eastern TR also does not fit easily into the latitude-dependent picture described above. There appears to be a connection between the two halves, with glacial flow from east TR down into SP (4), and also some shared color features with wisps of $CH_4$-rich material with colors similar to east TR extending westward into SP evident in Fig. 4G. $CH_4$ ice has a low density and could perhaps be transported as a crust on glacially flowing $N_2$ and CO ice.

Water ice presents a number of puzzles on Pluto. Its infrared spectral signature is associated with two very distinct shorter wavelengths color units. $H_2O$-rich outcrops in Virgil Fossa and Viking Terra (Fig. 2) show a distinct, reddish color in Fig. 4A, unlike the more neutral coloration of the $H_2O$-rich outcrops around Pulfrich crater. Rugged mountains such as Zheng-He and Norgay Montes expected to be composed of $H_2O$ ice show comparatively weak $H_2O$ spectral signatures.

Charon presents its own mysteries. The reddish polar region of Mordor Macula is a unique and striking feature not seen on other outer solar system icy satellites. The latitudinal dependence of its distribution suggests a mechanism involving seasonal cold trapping of volatiles such as $CH_4$ that would not otherwise be stable at Charon's surface. During Charon's long winter, polar latitudes remain unilluminated for multiple Earth decades, during which time they can cool to temperatures below 20 K (e.g., 55). Potential sources of $CH_4$ briefly resident in Charon's surface environment could be outgassing from Charon's interior and Pluto's escaping atmosphere, as discussed in (5,55). Seasonally cold-trapped $CH_4$ would be rapidly photolyzed by solar Ly-alpha radiation, roughly half of which arrives at Charon's surface indirectly via scattering by interplanetary hydrogen. Resulting radicals will combine into heavier products that are sufficiently non-volatile to remain after Charon's pole emerges back into the sunlight and warms to summer temperatures in the 50 to 60 K range. Further photolysis and radiolysis leads to production of reddish tholins as discussed above. This hypothesis predicts that Charon's southern hemisphere should exhibit a similar high-latitude reddish patch.

Charon's isolated ammonia-rich areas are also intriguing. $NH_3$ is a potentially important geochemical material in icy satellites that has heretofore mostly eluded detection via remote sensing techniques. A possible scenario for its appearance in just a few of Charon's craters is that these impacts dredged up the $NH_3$ from below Charon's surface too recently for it to have been destroyed by space weathering processes. It is also possible that a subset of impactors deliver $NH_3$-rich material, or that Charon's subsurface is heterogeneous, with local subsurface concentrations of $NH_3$ emplaced during an earlier era of cryovolcanic activity being subsequently exhumed by impacts.

# Conclusions

We have presented spatially-resolved visible and near infrared observations of the encounter hemispheres of Pluto and Charon, obtained by the New Horizons spacecraft on 2015 July 14. Data returned so far reveal complex spatial distributions of Pluto's $CH_4$, $N_2$ and CO ices as well as the local emergence of water ice bedrock and broad expanses of accumulated tholin at low latitudes. The data point to atmospheric and geological processes having acted over a range of timescales to create the currently observed surface. On Charon, the presence and distribution of localized ammonia ice outcrops and of reddish circumpolar material raise questions on the exogenous and endogenous processes acting on this large satellite.

Much of the data collected by New Horizons have yet to be transmitted back to Earth. They will enable us to quantitatively map the composition, state, and texture distributions of the system's inventory of materials in order to disentangle the history of these icy bodies and understand their place in the broader context of the outer solar system.

# Acknowledgments

This work was supported by NASA's New Horizons Project. S. Philippe and B. Schmitt acknowledge the Centre National d'Etudes Spatiales (CNES) for its financial support through its "Système Solaire" program. S.A. Stern is also affiliated with Florida Space Institute, Uwingu LLC, Golden Spike Co., and World View Enterprises. As contractually agreed to with NASA, fully calibrated New Horizons Pluto system data will be released via the NASA Planetary Data System at https://pds.nasa.gov/ in a series of stages in 2016 and 2017 as the data set is fully downlinked and calibrated.

# Figures and Captions

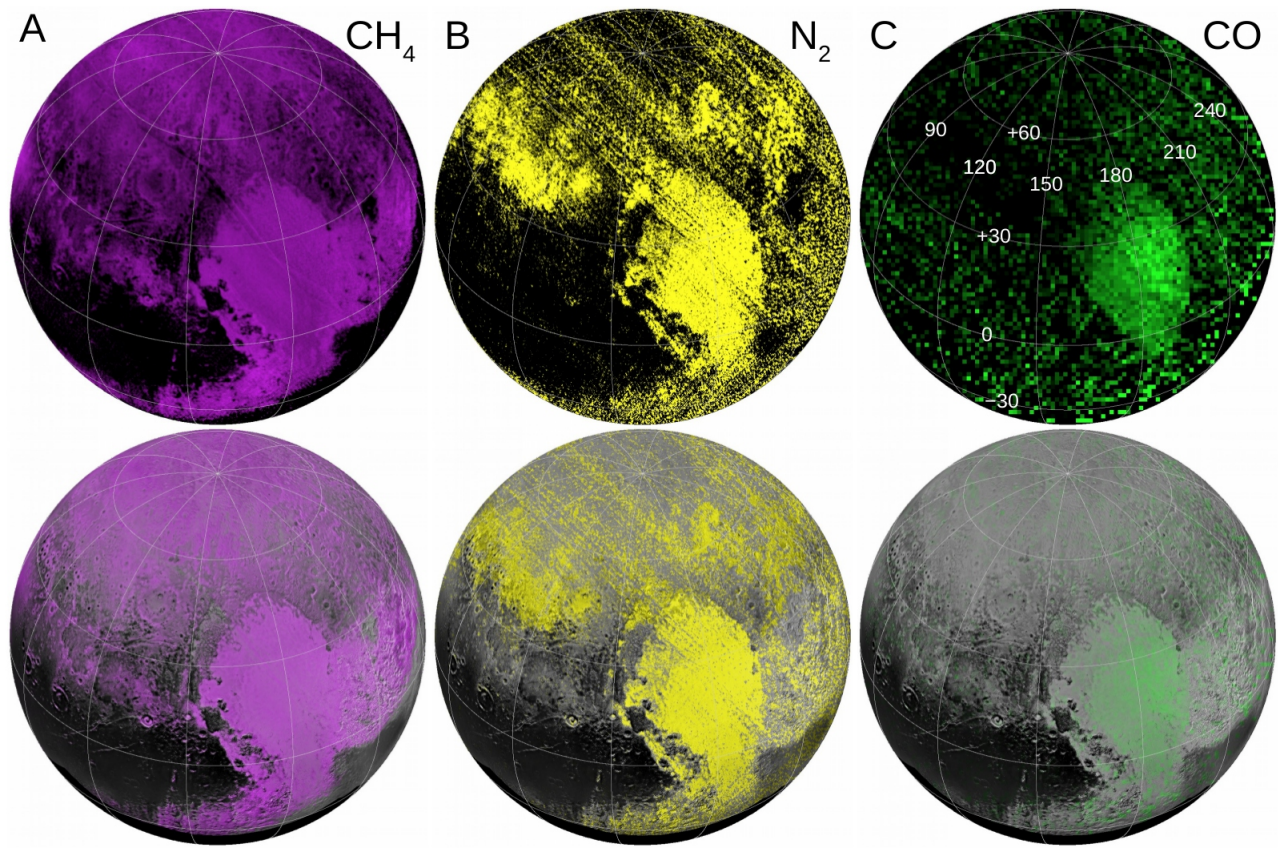

**Fig. 1. LEISA maps of Pluto's volatile ices $CH_4$, $N_2$, and CO.** For each species, the top panel shows the LEISA map, with brighter colors corresponding to more absorption, and the bottom panel shows it on a base map made from LORRI images reprojected to the geometry of the LEISA observation. The $CH_4$ absorption map (Panel A) shows the equivalent width of the 1.3 to 1.4 µm band complex. The $N_2$ absorption map (Panel B) is a ratio of the average over the band center (2.14 to 2.16 µm) to that of adjacent wavelengths (2.12 to 2.14 and 2.16 to 2.18 µm). The CO absorption map (Panel C) is based on a ratio, with the band center taken as wavelengths from 1.56 to 1.58 µm, and adjacent continuum wavelengths being 1.55 to 1.56 and 1.58 to 1.59 µm.

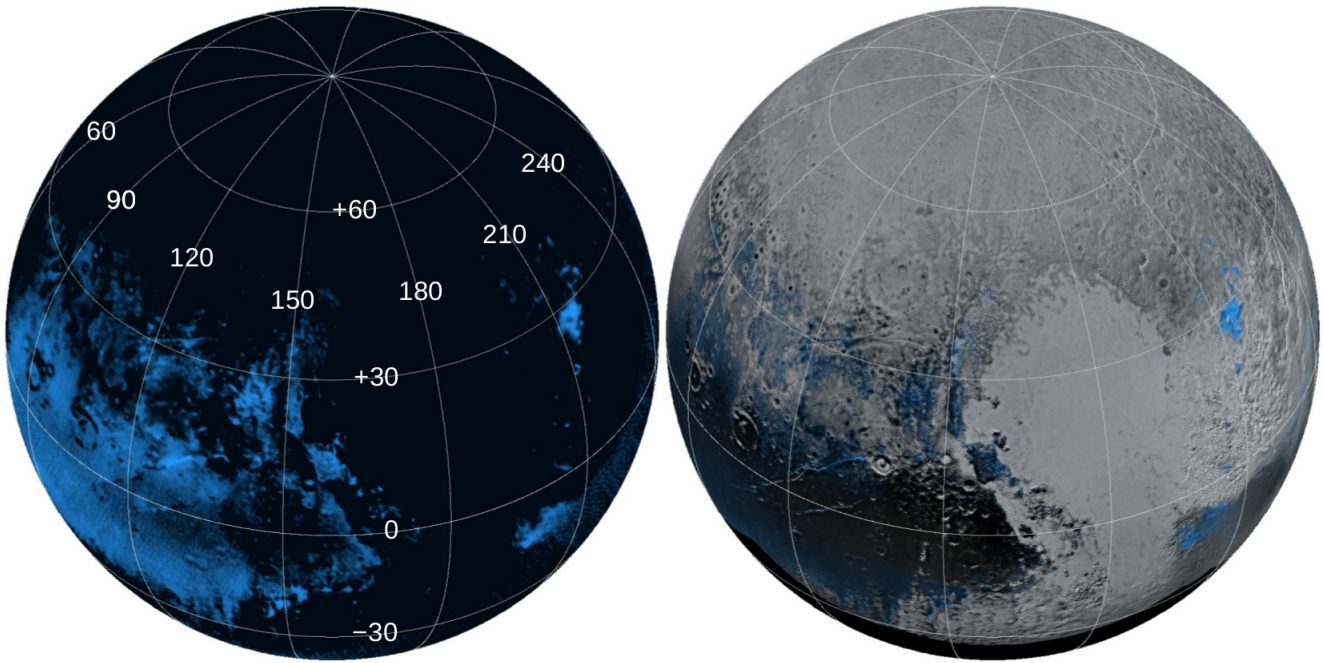

**Fig. 2. LEISA map of Pluto's non-volatile H$_2$O ice.** This map (left) shows the correlation coefficient between each LEISA spectrum and a template Charon-like H$_2$O ice spectrum (e.g., 47,49), highlighting where H$_2$O absorption is least contaminated by other spectral features. The LEISA map is superposed on the reprojected LORRI base map at right.

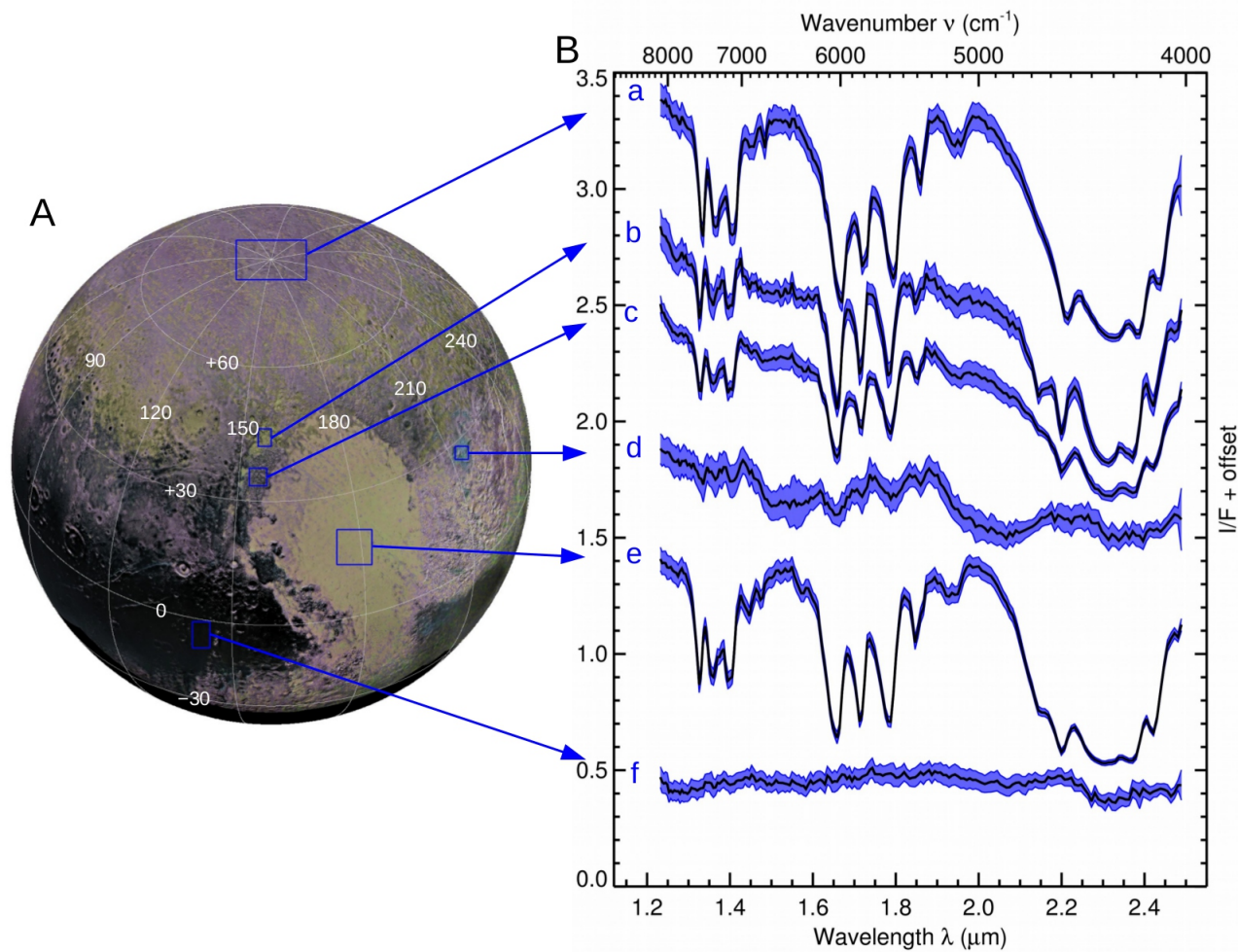

**Fig. 3. LEISA spectra of Pluto.** Panel A shows a context map produced by averaging the red, green, and blue values from each of the four colored maps across the bottom row in Fig. 1 and the right panel in Fig. 2. Specific intensity (*I/F*) spectra averaged over regions in blue boxes are plotted in Panel B, with envelopes indicating the standard deviations within the boxes. These regions were selected to highlight Pluto's spectral diversity. Vertical offsets for spectra 'a' through 'f' are +2.3, +1.7, +1.5, +1.2, +0.45, and 0, respectively. Pluto's north pole ('a') shows strong absorptions by $CH_4$ ice. Spectrum 'b' is a region characterized by a strong $N_2$ ice absorption at 2.15 µm and weak $H_2O$ ice bands at 1.5 and 2 µm. Spectrum 'c' is al-Idrisi Montes, very similar to 'b', except without the $N_2$ absorption. The area around Pulfrich crater ('d') has $H_2O$ ice absorptions at 1.5, 1.65, and 2 µm and comparatively weak $CH_4$ ice absorptions. Spectrum 'e' is the center of Sputnik Planum, with strong $CH_4$ bands, and also the $N_2$ ice absorption at 2.15 µm and CO ice absorption at 1.58 µm. Spectrum 'f' is eastern Cthulhu Regio, with weak $H_2O$ ice absorptions at 1.5 and 2 µm and a feature attributed to heavier hydrocarbons at 2.3 µm.

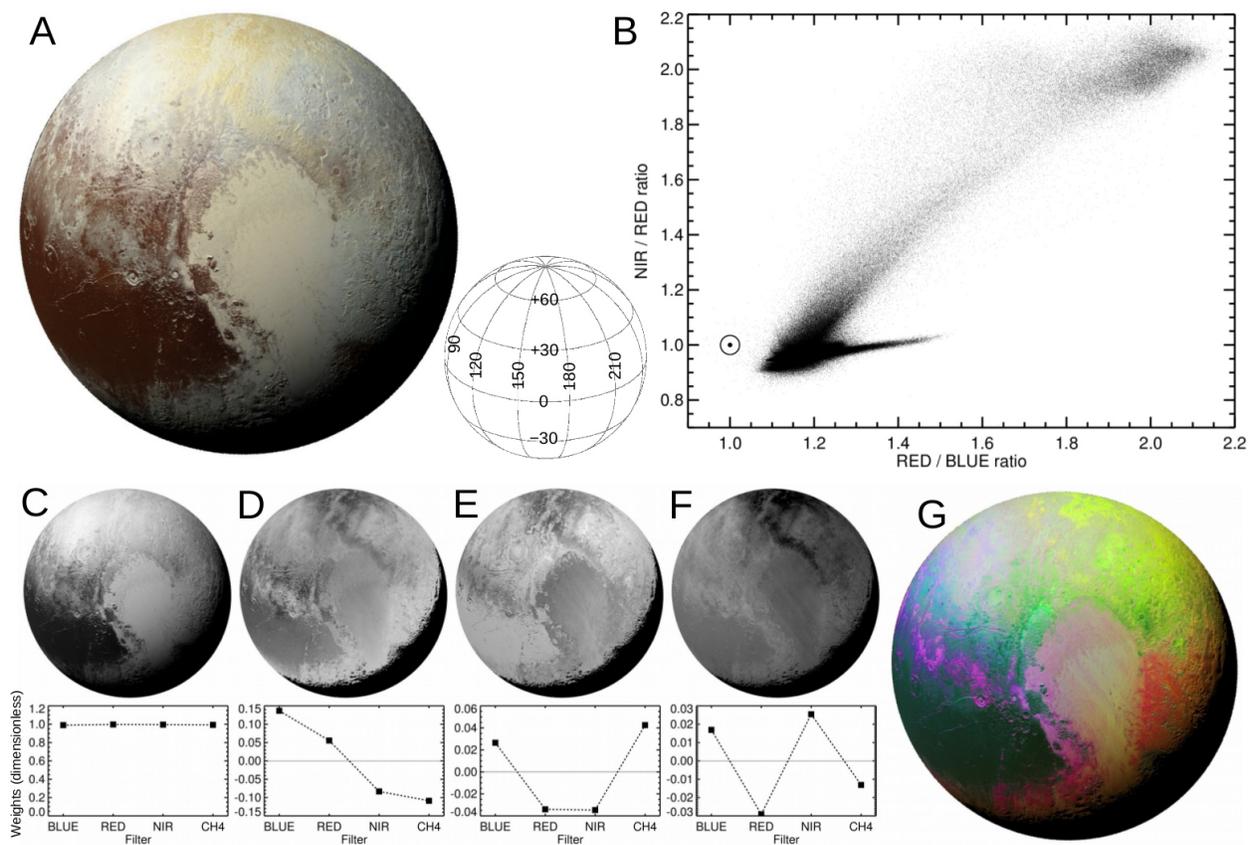

**Fig. 4. MVIC colors of Pluto.** Panel A shows "enhanced" color with MVIC's BLUE, RED, and NIR filter images displayed in blue, green, and red color channels, respectively. Geometry is indicated by the wire grid. Panel B shows the distribution of NIR/RED and RED/BLUE color ratios, excluding regions where the incidence angle from the sun or the emission angle to the spacecraft exceeds 70° from the zenith. The sun symbol indicated neutral colors and redder colors extend up and to the right. Panels C, D, E, and F show principal component images and eigenvectors for PC1 through PC4, respectively. Panel G is a false color view with shading from PC1 and the hue set by PC2, PC3, and PC4 being displayed in red, green, and blue channels, respectively.

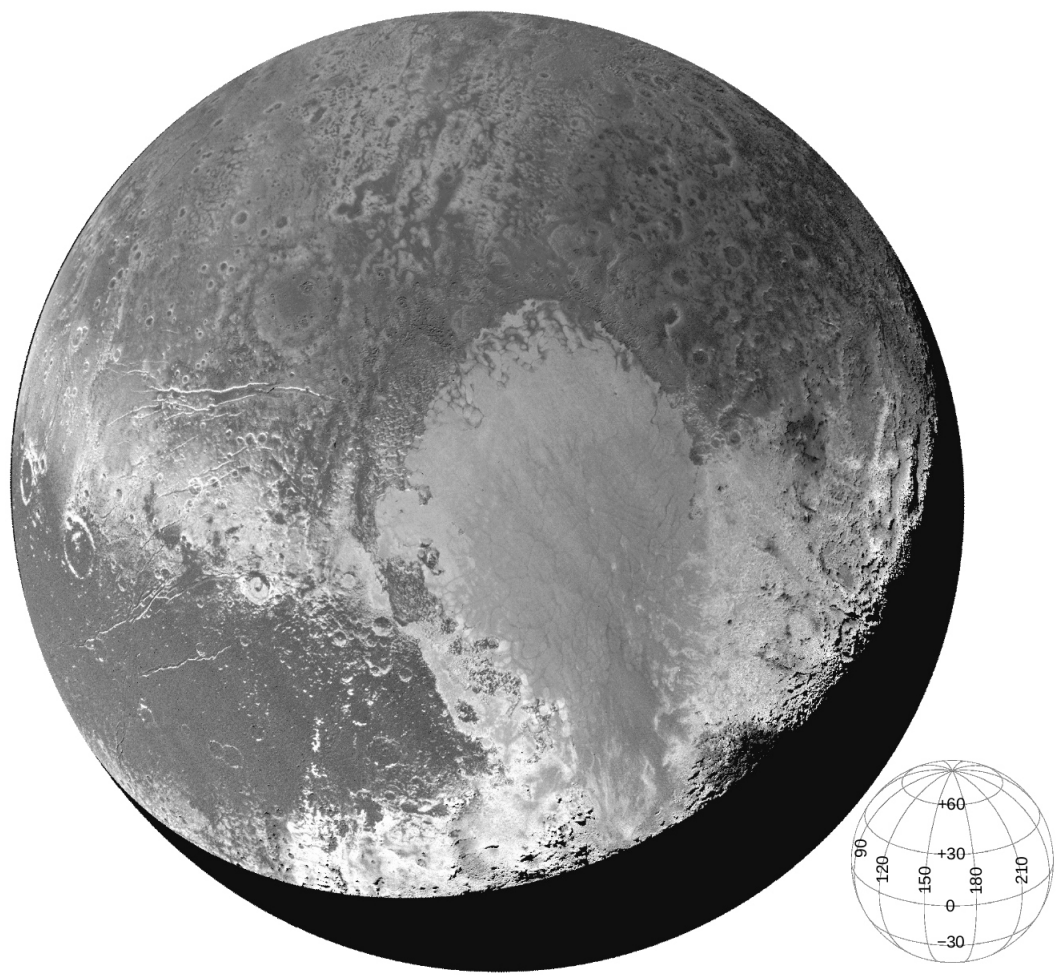

**Fig. 5. Pluto MVIC CH$_4$ absorption map.** The equivalent width of absorption in the MVIC CH4 filter, is computed by comparison with the NIR and RED filter images (see supplementary text for details). This filter is centered on a weaker CH$_4$ ice absorption than the one mapped with LEISA data in Fig. 1. Brighter shades correspond to stronger CH$_4$ ice absorption. Differences between the maps are discussed in the text. Except for a sliver of poorly illuminated terrain along terminator where geometric effects become extreme, most of the contrast in this map traces regional variations in CH$_4$ ice absorption.

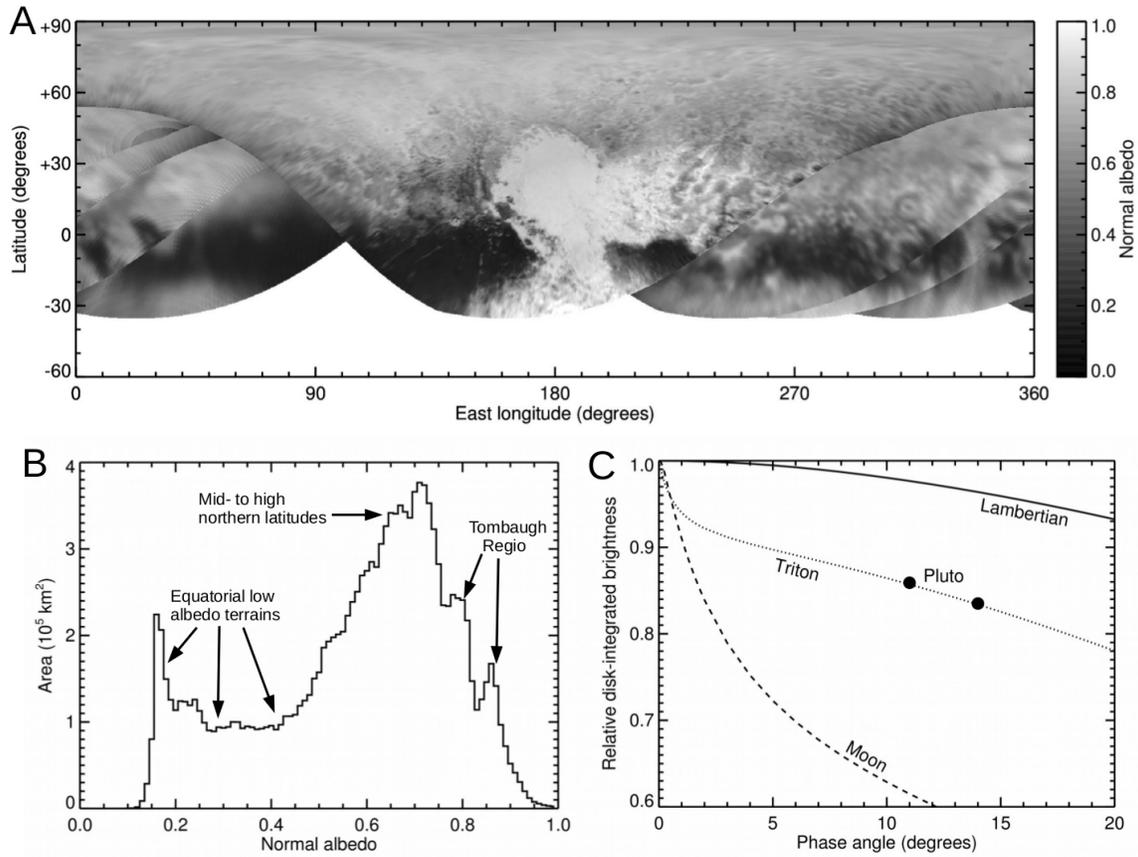

**Fig. 6. LORRI albedos on Pluto.** Panel A shows normal albedo across Pluto from LORRI panchromatic images sensitive to wavelengths from 350 to 850 nm (10). Panel B shows a histogram of albedo values. Panel C compares the decline of Pluto's disk-integrated brightness with phase angle with a Lambertian sphere, the icy satellite Triton, and Earth's Moon. The Triton curve is based on Voyager green filter and ground-based observations with an effective wavelength of 550 nm (40). The lunar curve is for Johnson *V* filter wavelength of 550 nm. The Pluto points show LORRI brightness after correcting for lightcurve variability, relative to zero phase data from ground-based monitoring (43, Bessel *R* filter, 630 nm).

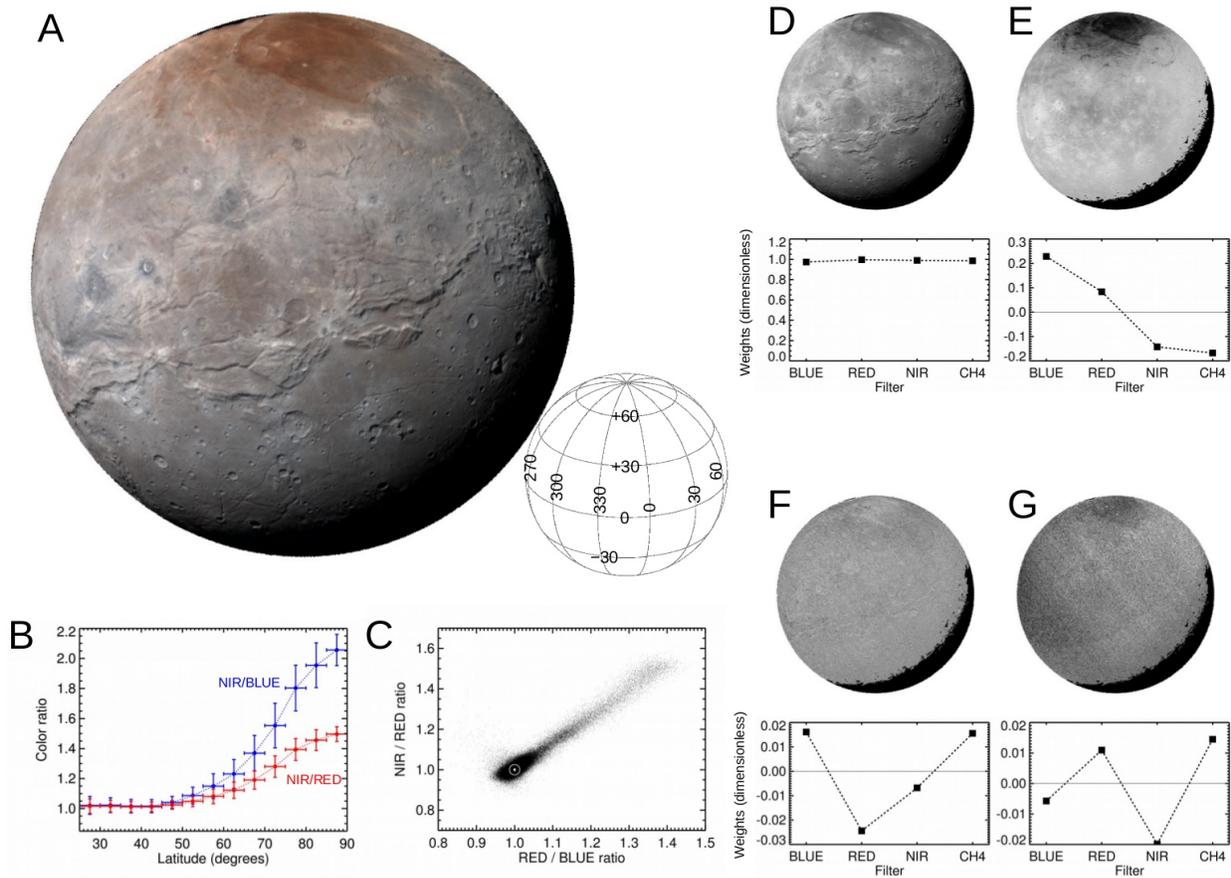

**Fig. 7. MVIC colors of Charon.** Panel A shows enhanced color, with MVIC's BLUE, RED, and NIR filter images displayed in blue, green, and red color channels, respectively. Geometry is indicated by the associated wire grid. Panel B shows MVIC NIR/BLUE and NIR/RED color ratio means and standard deviations averaged over 5° latitude annuli, excluding points near the limb with emission angles greater than 75°. Panel C shows a scatterplot of NIR/RED and RED/BLUE color ratios, excluding incidence and emission angles exceeding 70°. Most pixels cluster near solar colors, with a mixing line extending toward redder colors at upper right. Panels D, E, F, and G show the principal component images and eigenvectors for principal components 1, 2, 3, and 4, respectively.

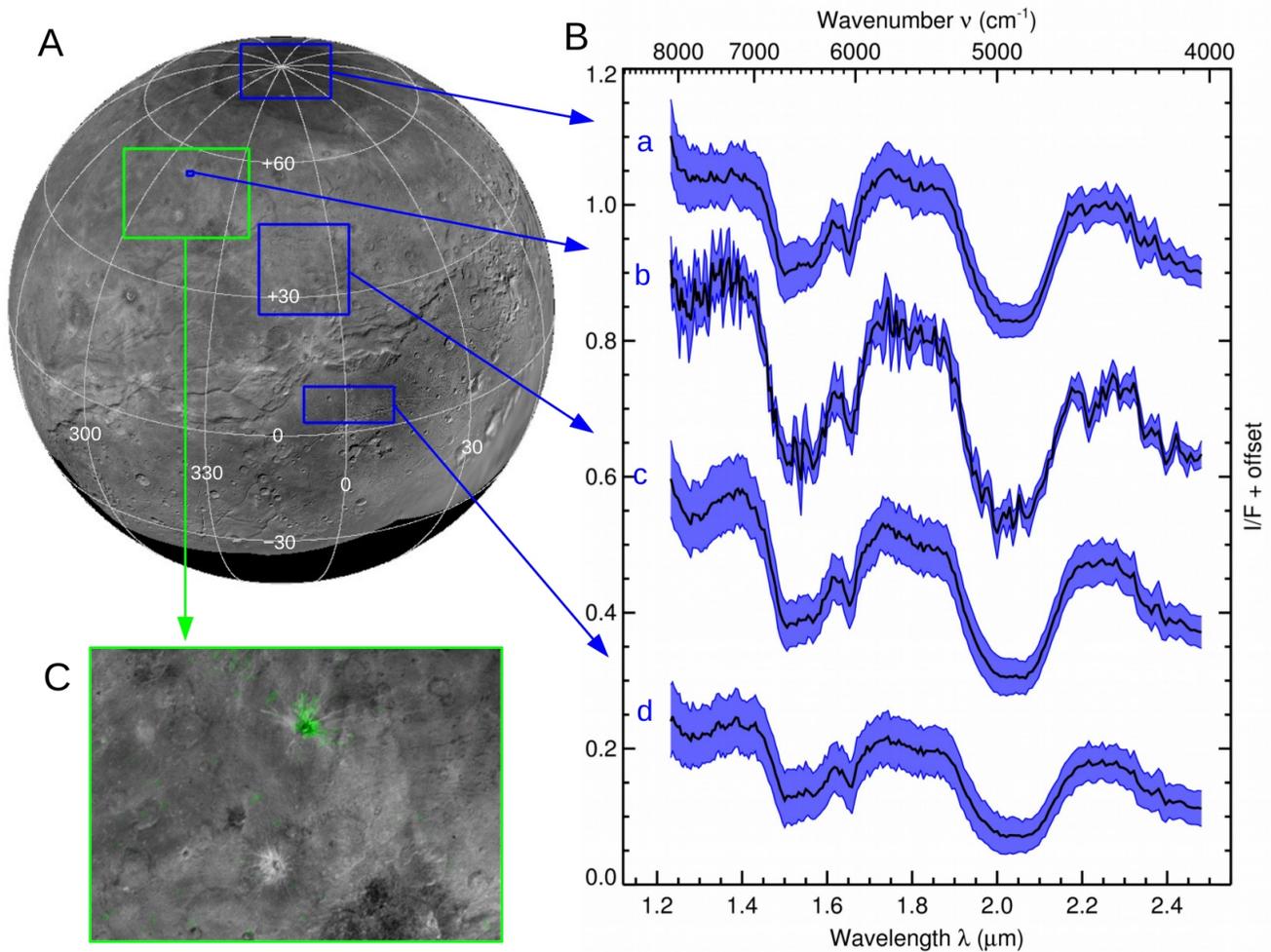

**Fig. 8. LEISA spectra of Charon.** Panel A shows a LORRI composite base map with blue rectangles indicating regions where I/F spectra were averaged for plotting in Panel B. Vertical offsets for spectra 'a' through 'd' are +0.7, +0.4, +0.2, and 0, respectively. All four spectra show the characteristic absorption bands of cold, crystalline $H_2O$ ice, at 1.5, 1.65 and 2 µm. Charon's north pole ('a') shows a little more continuum absorption towards short wavelengths, but no other obvious differences relative to spectra from lower latitudes on Charon. Spectrum 'b' is a region around Organa crater showing $NH_3$ absorption at 2.22 µm. Panel C zooms into the region indicated by the green box in Panel A, with 2.22 µm absorption mapped in green to show the spatial distribution of $NH_3$-rich material (Fig. S6 shows the full map). Spectra 'c' and 'd' compare plains units above and below the tectonic belt.

# Supplementary Materials

www.sciencemag.org

Supplementary text

Figs. S1, S2, S3, S4, S5, S6

Reference (56)

# Supplementary Materials for

## Surface Compositions Across Pluto and Charon

W.M. Grundy, R.P. Binzel, B.J. Buratti, J.C. Cook, D.P. Cruikshank, C.M. Dalle Ore, A.M. Earle, K. Ennico, C.J.A. Howett, A.W. Lunsford, C.B. Olkin, A.H. Parker, S. Philippe, S. Protopapa, D.C. Reuter, B. Schmitt, K.N. Singer, A.J. Verbiscer, R.A. Beyer, M.W. Buie, A.F. Cheng, D.E. Jennings, I.R. Linscott, J.Wm. Parker, P.M. Schenk, J.R. Spencer, J.A. Stansberry, S.A. Stern, H.B. Throop, C.C.C. Tsang, H.A. Weaver, L.A. Young, and the New Horizons Science Team.

correspondence to:  w.grundy@lowell.edu

**This PDF file includes:**

Supplementary Text
Figs. S1 to S6

## Supplementary Text

MVIC radiometric calibration

MVIC throughput and I/F calibration were achieved through a combination of cruise stellar calibrations for absolute throughput of the RED channel followed by a channel-by-channel relative calibration using the global photometry of Charon as a calibration standard, matched to the global photometry of Charon derived from HST ACS HRC F435W and F555W observations (56). New Horizons observations used for calibration were geometrically corrected to re-weight the contribution of Charon's red polar spot given the sub-observer latitude of the HST observations. In addition to calibrating the system throughput, an additional instrument calibration was performed on the data presented in this paper: on approach, the gain of the NIR channel was found to change from scan to scan in a non-predictable fashion on one of the instrument's two power sides. The gain remained constant during each scan. The problematic power side was used for the P_COLOR2 Pluto observation described in this paper. In order to correct for this drift, earlier overlapping images taken on the alternate power side were used to bootstrap a gain correction. Sputnik Planum was used as the control region for this bootstrapped correction, for its nearly neutral color and relative lack of albedo contrasts. The PC_MULTI_MAP_B_17 observation obtained 2015 July 13 at 3:38 UT, at a scale of 32 km/pixel was used as the control data.

MVIC color $CH_4$ equivalent width

MVIC's CH4 filter is centered on the strongest $CH_4$ ice absorption band in MVIC's wavelength range, at 890 nm. We estimate the equivalent width of absorption in that band from MVIC's RED, NIR, and CH4 bands as follows. We start by forward modeling the parameter space of possible equivalent widths and spectral slopes, assuming reflectance is a linear function in wavelength with the addition of a perfect absorption band (zero reflectance) centered at 890 nm. This simple model is multiplied by a solar spectrum and sampled according to MVIC's wavelength-dependent filter transmissions, throughputs, and quantum efficiencies to compute a grid of synthetic CH4/NIR and RED/NIR ratio values as functions of slope and equivalent width. Equivalent width and slope maps can then be computed from actual MVIC CH4/NIR and RED/NIR ratio images, pixel by pixel, by interpolating the forward-modeled grids to retrieve the corresponding equivalent width and slope values. The equivalent width map is shown in Fig. 5 while the slope map is shown in Fig. S3.

Pluto normal albedo map

To map albedos or reflectances across a planet's surface, a photometric model is needed to account for changes in viewing and illumination geometry across the field of view of individual images and across multiple observations obtained at different times. We used this simple photometric model from (41):

$$I/F = A f(\alpha) \frac{\cos(i)}{\cos(i)+\cos(e)} + (1-A)\cos(i)$$

where $I/F$ is the specific intensity, $i$ is the incident angle, $e$ is the emission angle, and $f(\alpha)$ is the surface solar phase function, which includes changes in intensity due to the physical character of the surface (roughness, the single scattering albedo, the single particle phase function, the compaction state of the optically active portion of the regolith, and coherent backscatter). The first term describes singly scattered radiation while the second term describes multiple scattering;

*A* is a parameter that gives the fraction of each component. Fitting this function to the LORRI observations of Pluto's surface gives *A* = 0.7, in which 30% of the reflected photons are multiply scattered. This function is similar to those found for icy moons of Saturn (41). Because the images from New Horizons were obtained at solar phase angles of 11° and above, it is necessary to use ground-based observations to correct I/F to normal reflectance, which is the albedo for incident, emission, and solar phase angles all equal to 0°. To correct to 0° we used the phase behavior from ground-based observations below 2° (42,43) and full-disk New Horizons LORRI images at 11° and 14°.

Charon's $NH_3$ absorption

$NH_3$ ice has characteristic absorption bands at 2.00 and 2.22 µm (the exact wavelength ranges from 2.20 to 2.24 µm depending on the hydration state). The 2.00 µm band is hard to discern on Charon since the ubiquitous $H_2O$ ice also absorbs strongly at that wavelength, but the 2.22 µm $NH_3$ band coincides with an $H_2O$ continuum region. We mapped this band in LEISA data by computing *I/F* averages over wavelengths from 2.20 to 2.24 µm, covering the $NH_3$ ice absorption band, and also over adjacent continuum wavelengths from 2.10 to 2.17 µm and from 2.26 to 2.29 µm. Dividing the continuum average image by the band average image gives larger values where the $NH_3$ absorption is stronger. The resulting $NH_3$ absorption map is shown in Fig. S6. It is spatially fairly uniform, with only a few believable features rising above the noise, the most prominent of which corresponds to Organa crater at 310.9° E, 54.3° N. The region around Organa crater is enlarged in Fig. 8C.

**Fig. S1.**

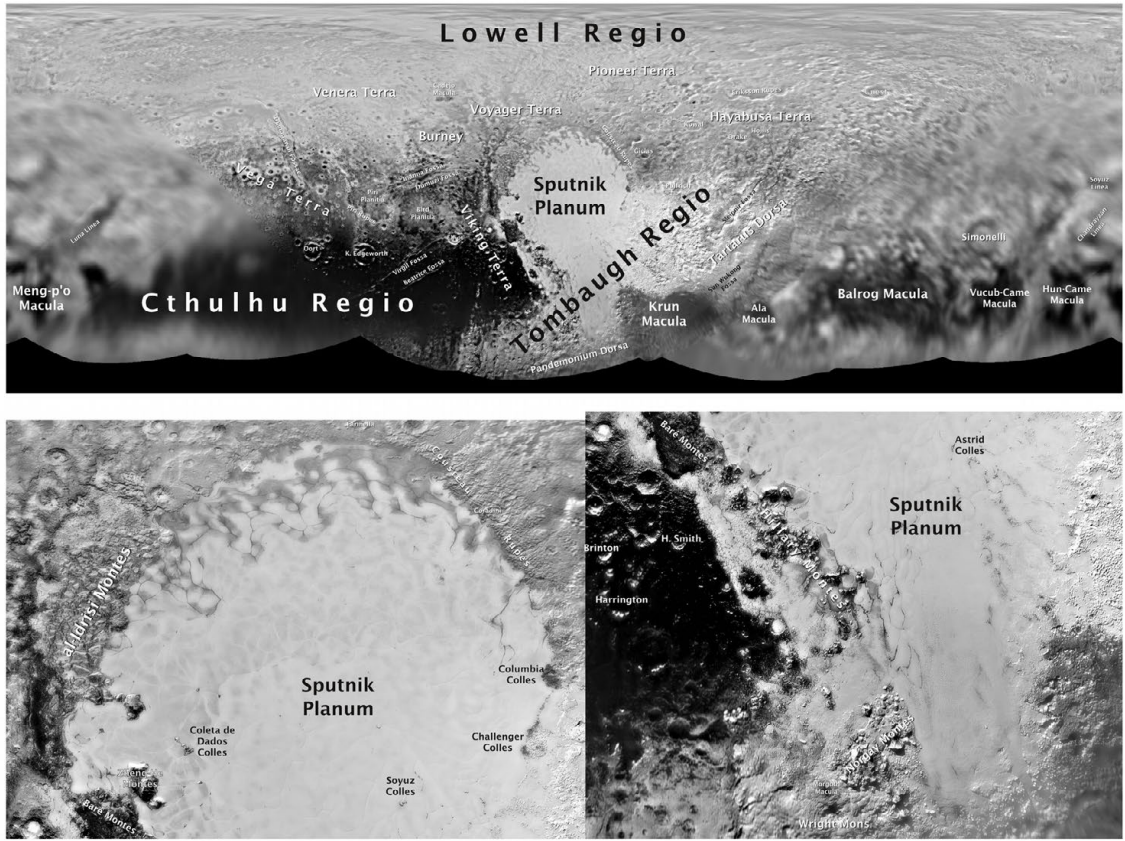

**Nomenclature maps for Pluto.** This figure is duplicated from Moore et al. (4). All names are informal.

**Fig. S2**

**Nomenclature map for Charon.** This figure is duplicated from Moore et al. (4). All names are informal.

**Fig. S3**

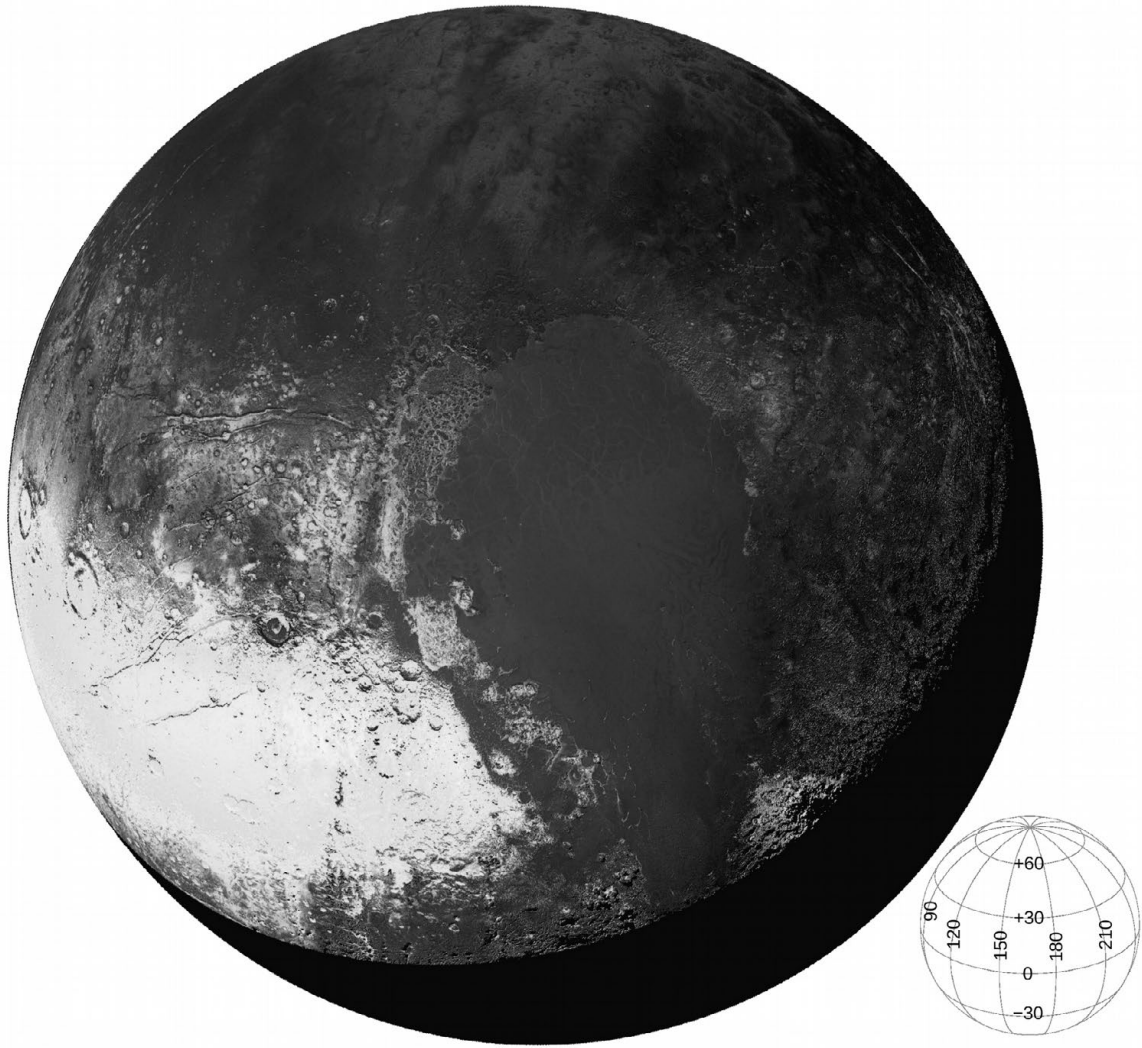

**Pluto spectral slope.** This map is a computed in concert with the 890 nm equivalent width map shown in Fig. 5. Brighter areas correspond to redder spectral slopes over the 540 to 975 nm wavelength range sampled by MVIC's RED, NIR, and CH4 filters.

**Fig. S4**

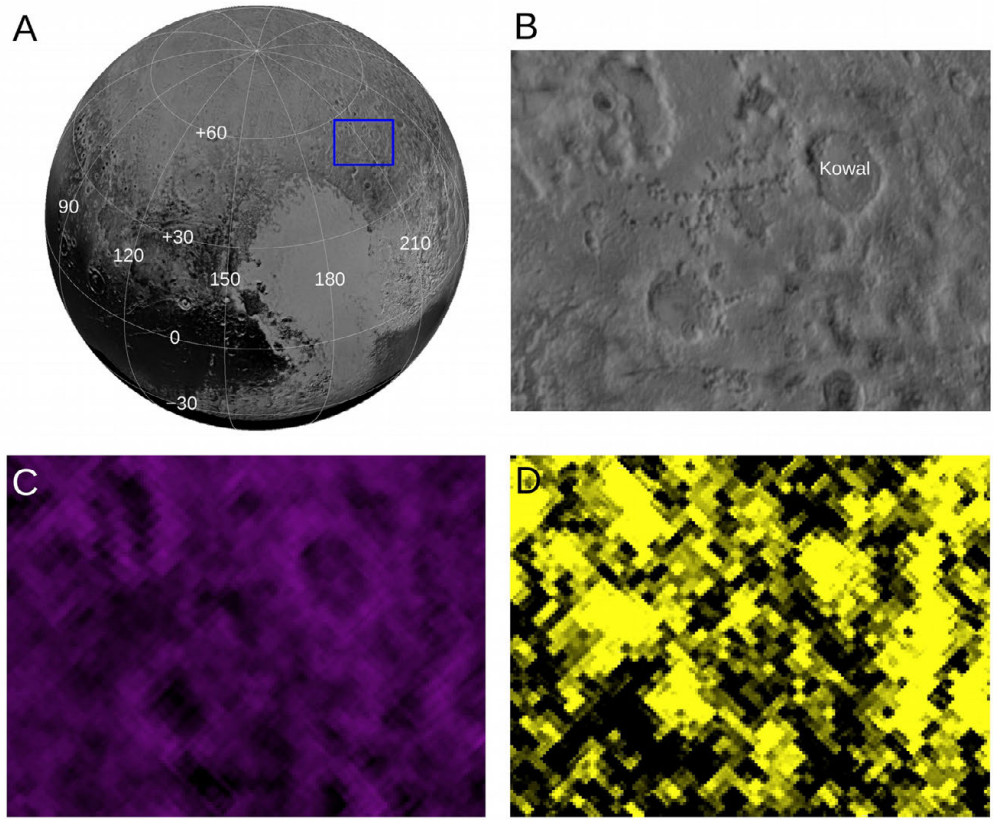

**Pluto's contrasting CH₄ and N₂ ice distributions.** Panel A shows a LORRI base map for context, with a blue box indicating a region to be enlarged in the vicinity of Kowal crater. Panels B, C, and D zoom in on the corresponding area of the base map, the CH$_4$ map, and the N$_2$ map, respectively. At these latitudes, CH$_4$ ice absorption tends to be associated with ridges and crater rims, while N$_2$ ice absorption appears more prominent on crater floors. CH$_4$ ice accumulating on local topographic highs could be related to construction of edifices like the bladed terrain in Tartarus Dorsa.

**Fig. S5**

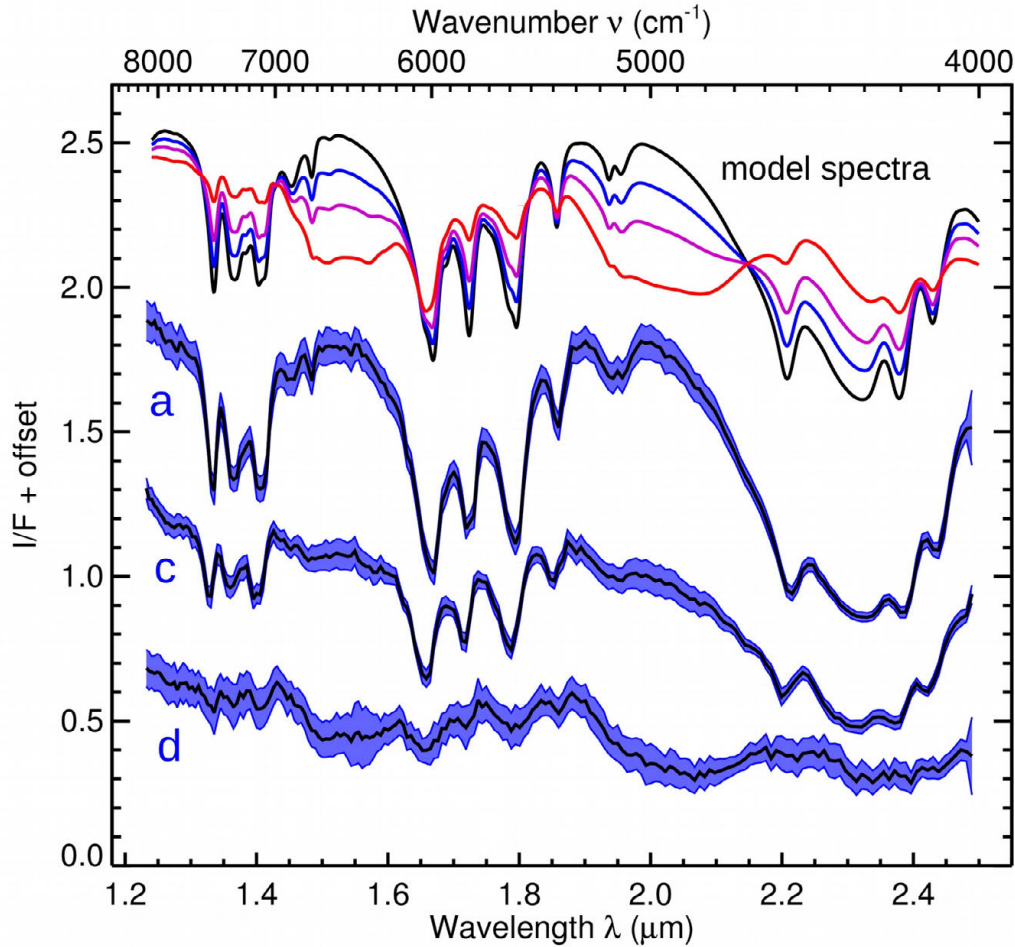

**Models of CH$_4$ plus H$_2$O compared with Pluto spectra.** Pluto's near-infrared spectrum is dominated by the numerous strong absorption bands of CH$_4$ ice, making it difficult to detect absorbers with broad absorption features, such as H$_2$O ice. The H$_2$O correlation map in Fig. 2 highlights regions with the most conspicuous H$_2$O absorption bands, but many other areas show more subtle influence of H$_2$O absorptions via reduced albedos at 1.5 and 2 µm. These more subtle effects of H$_2$O absorption are illustrated with model spectra at the top of the plot, showing the influence of adding H$_2$O ice to a terrain that is spectrally dominated by CH$_4$ ice. The black model curve is for pure CH$_4$ ice. The colored model curves include various amounts of H$_2$O ice, ranging from the blue one having the least (20% H$_2$O ice in an areal mixture) to the red one having the most (70% H$_2$O in an areal mixture). Pluto spectra 'a', 'c', and 'd' are duplicated from Fig. 3B, corresponding to Lowell Regio, al-Idrisi Montes, and the H$_2$O-rich region around Pulfrich crater, with vertical offsets of +0.8, +0.3, and 0, respectively. The model spectra were offset by +1.6.

**Fig. S6**

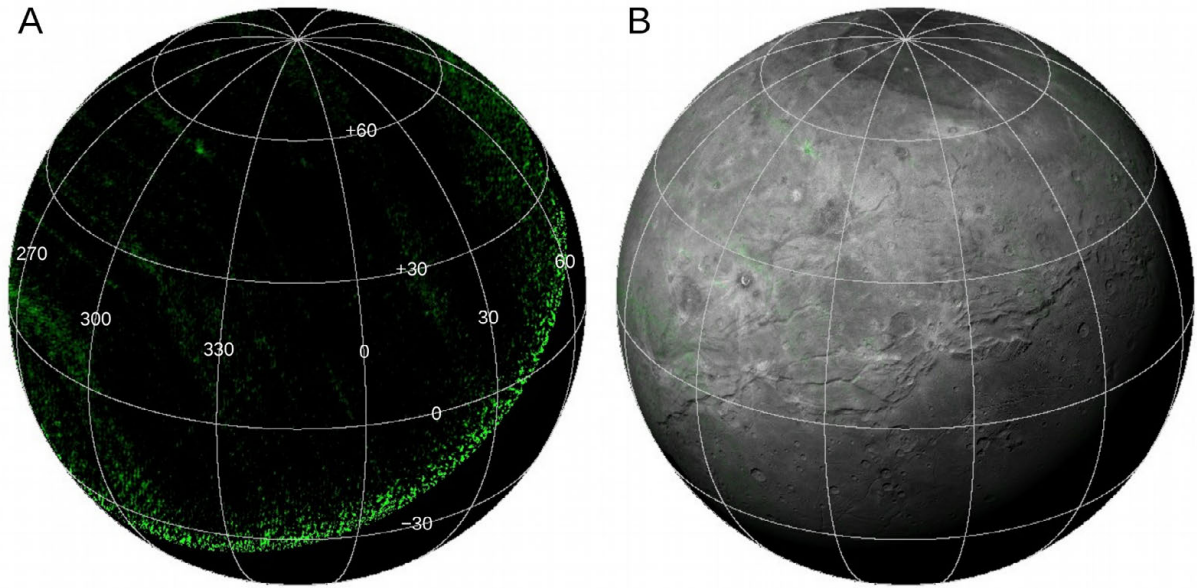

**Charon ammonia absorption map.** This map highlights Charon's 2.22 µm $NH_3$ absorption band. The map is shown without accoutrements at left and coloring the reprojected LORRI base map at right.